\documentclass[a4paper,12pt]{article}
\usepackage[utf8]{inputenc}
\usepackage{amsmath}
\usepackage{amsfonts}
\usepackage{amssymb}

\usepackage[T1]{fontenc}
\usepackage{gensymb}
\usepackage[hidelinks]{hyperref}
\usepackage{lmodern}
\usepackage{algorithm}
\usepackage{algpseudocode}

\usepackage{graphicx}
\usepackage[
backend=biber,
style=numeric,
sorting=anyt
]{biblatex}
\usepackage[left=2cm,right=2cm,top=2cm,bottom=2cm]{geometry}

\author{AIT HADDOU Marwan \\ \small{\url{marwan.aithaddou@edu.uca.ac.ma}}}
\date{\small{\today} \\ }
\title{Quasi-normal modes of near-extremal black holes in dRGT massive gravity using Physics-Informed Neural Networks (PINNs)}
\begin{document}

\maketitle
\abstract{In this study, we demonstrate the use of physics-informed neural networks (PINNs) for computing the quasinormal modes (QNMs) of black holes in de Rham-Gabadadze-Tolley (dRGT) massive gravity. These modes describe the oscillation frequencies of perturbed black holes and are important in understanding the behavior of these objects. We show that by carefully selecting the hyperparameters of the PINN, including the network architecture and the training data, it is possible to achieve good agreement between the computed QNMs and the approximate analytical formula in the near-extremal limit for the smallest mode number. Our results demonstrate the effectiveness of PINNs for solving inverse problems in the context of QNMs and highlight the potential of these algorithms for providing valuable insights into the behavior of black holes.}

\section{Introduction}

Quasinormal modes (QNMs) are complex numbers that describe the oscillation modes of a black hole that has been perturbed. They play a crucial role in our understanding of the behavior of black holes. The real part of QNMs corresponds to the oscillation frequency of the black hole, while the imaginary part corresponds to the damping rate of the oscillation. The calculation of QNMs is a difficult task, as it involves solving a differential equation with appropriate boundary conditions in a highly complex spacetime.

In recent years, Physics-informed neural networks (PINNs) have emerged as a machine learning algorithm that excels at tackling complex physics problems. These algorithms are especially useful for solving inverse problems, where the goal is to determine the unknown parameters of a system based on observed data. PINNs stand out due to their unique capability of incorporating physical constraints and laws directly into the neural network architecture. By doing so, they learn the solution to a problem by accounting for the underlying physics, rather than just fitting to a set of data points.

PINNs have been increasingly utilized to compute Quasinormal modes (QNMs) of black holes in recent times. When the black hole is near-extremal, the effective potential of the radial wave equation for QNMs can be expressed in terms of the well-known Poschl-Teller potential, for which an exact solution exists. As a result, the QNMs problem can be seen as an inverse problem that can be solved using PINNs. By utilizing the known solution of the Poschl-Teller potential to train a PINN and approximate the QNMs \cite{Berti2009TOPICALRQ}, highly precise approximations of the QNMs can be obtained for a broad range of black hole spacetimes.

Selecting the right hyperparameters can be a major challenge when utilizing PINNs to calculate QNMs. These parameters, such as the network architecture, activation functions, and training data, can have a significant impact on the performance of the algorithm. Therefore, it is crucial to carefully choose these parameters in order to achieve favorable outcomes.

In conclusion, PINNs have emerged as a promising method for calculating QNMs of black holes and can offer valuable insights into the behavior of these entities. With the progress of the machine learning field, PINNs are expected to play an increasingly vital role in tackling complex problems in astrophysics and other areas of physics.

\section{Computing QNMs with Physics-Informed Neural Networks}
Deep learning is a powerful machine learning technique that uses input data to generate predictions. This method identifies patterns in the data and utilizes them to minimize the difference between the predicted and actual output, while also enabling generalization to new inputs. One of the most popular types of deep learning algorithms is neural networks, which analyze the correlation between input and output data.

A neural network is composed of various layers, including an input layer, hidden layers, and an output layer. The input layer collects data, the hidden layers adjust weights and parameters to recognize patterns within the data, and the output layer generates predictions based on these patterns.

Physics-informed neural networks (PINNs) are a specific type of deep learning algorithm that incorporates the physical principles of the problem, which are often represented by nonlinear partial differential equations. In 2017, Raissi et al. were the first to suggest using PINNs to solve such equations.

To compute quasinormal frequencies (QNFs), we can use the DeepXDE library, which is designed for PINN applications. DeepXDE simplifies the process by providing built-in modules for the computational domain, PDE equations, boundary and initial conditions, constraints, training data, neural network architecture, and training hyperparameters. The steps to solve differential equations using DeepXDE are as follows:

\begin{algorithm}
\caption{The process for building PINN models in DeepXDE}\label{alg:cap}
\begin{algorithmic}[1]
\State  The physical equations that drive the problem are all set and passed to \texttt{deepxde.data...(...)} as parameters.
\State  Construct a neural network using the \texttt{deepxde.nn...(...)} module.
\State The PINN model is a combination of the neural network and physical constraints defined as parameters to \texttt{deepxde.Model(...)}.
\State The \texttt{Model.Compile(...)} function is used to define training parameters such as the specific choice for the optimiser.
\State  The PINN model can then be run using \texttt{Model.train(...)} for a defined number of training epochs.
\State Call \texttt{Model.predict(...)} to predict the solution.

\end{algorithmic}
\end{algorithm}

 \begin{figure}[h]
    \centering
    \includegraphics[width=0.8\textwidth]{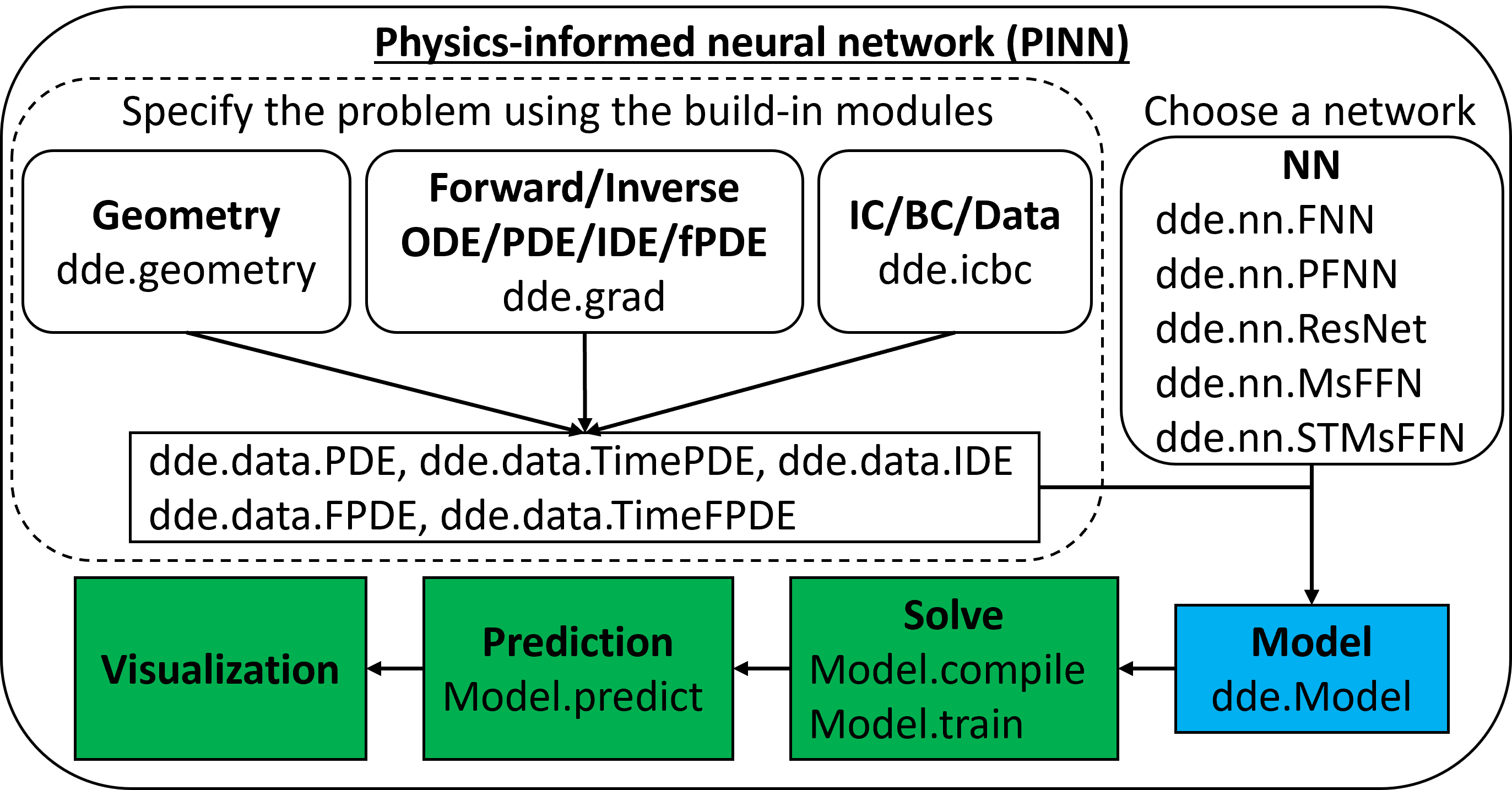}
    \caption{The PINNs algorithm in DeepXDE \cite{deepXDE}}
    \label{fig:algo}
\end{figure}
\newpage
\section{Quasi-normal modes of scalar perturbations in a pure de Sitter space}
\subsection{System Setup}
The static metric of a four-dimensional de Sitter (dS) space is given by:
\begin{equation}
ds^2 = -f(r) dt^2 + f^{-1}(r)^2 + r^2 d\Omega^2
\end{equation}
Here, $f(r) = 1- r^2/z^2$, where $z$ is the minimal radius of the dS space, and $r^2 d\Omega^2$ represents the metric on the two-dimensional sphere $S^2$ of radius $r$.

For a massive scalar field $\Psi$ that satisfies the Klein-Gordon equation:
\begin{equation}
\Psi^{;\nu} {}{;\nu} = m\Psi
\end{equation}
$\Psi$ can be separated as:
\begin{equation*}
\Psi = \dfrac{u}{r} e^{-\omega t} Y{l}(\Omega)
\end{equation*}
where $Y$ is the spherical harmonic function. By using the tortoise coordinate $dx = dr/f(r) = z tanh^{-1}(r/z)$, we can express the radial part in a Schrödinger-like equation:
\begin{equation}
\dfrac{d^2 u}{dx^2} + (\omega^2 - V(x)) u = 0
\end{equation}
where the effective potential is given by:
\begin{equation}
V(x) = -\dfrac{2 - m^2 z^2}{z^2 cosh^2(x/z)} + \dfrac{l(l+1)}{z^2 sinh^2(x/z)}
\end{equation}
By introducing a new variable $a = 1/ cosh^2
(x/l)$, equation (3) can be written as:
\begin{equation}
a(1-a)u^{''} + (1-\dfrac{3}{2}a)u^{'} + \dfrac{1}{4}(\dfrac{\omega^2 z}{a} - \dfrac{l(l+1)}{1-a} + 2 - m^2 z^2)u = 0
\end{equation}
The exact solution of this equation is given by:
\begin{equation}
u(a) \propto a^{\dfrac{-i\omega l}{2}}(1-a)^{\dfrac{l+1}{2}} {}_2F_1 (b - c + 1,d - c + 1, 2 - c,a)
\end{equation}
where:
\begin{align*}
b &= \alpha + \beta + \dfrac{1}{4}(1 + \sqrt{1 + 4 (2 - m^2 z^2)}),\\
c &= 2\alpha + 1,\\
b &= \alpha + \beta + \dfrac{1}{4}(1 + \sqrt{1 - 4 (2 - m^2 z^2)}),\\
\alpha &=-j\omega z /2
\end{align*}
The corresponding quasi-normal frequencies are:
\begin{equation}
\omega = -j\dfrac{1}{2}(2n + l + \dfrac{3}{2})
\end{equation}
The results were generated using the following hyperparameters: two fully connected neural networks ($PFNN$) with a depth of 4 (i.e., 3 hidden layers) and the structure $[1, [20, 20], [20, 20], [20, 20], 2]$ were used. The non-linear activation function $tanh$ was employed, along with $Adam$ optimisers with a learning rate of 0.0001 and 160,000 training epochs. The training data consisted of 250 domain points, and the dataset used for training contained 250 actual values of the Quasi-normal Modes (QNMs) ($\Psi(y)$) uniformly distributed in the domain [1.1, 2.5] for x.

\begin{table}[h]
\centering
\begin{tabular}{|c|c|c|c|c|c|}
\hline
\textbf{(n, l)} & \textbf{Ref \cite{Du:2004jt}} & \textbf{FNN \cite{Ovgun:2019yor}} & \textbf{PINN} & \textbf{PE (FNN) \cite{Ovgun:2019yor}} & \textbf{PE (PINN)} \\
\hline
(0, 0) & 0 - 3000j & 0 - 3017j & 0.0 - 2997.10j & 0.56 & 0.0965 \\
\hline
(0, 1) & 0 - 4000j & 0 - 3994j & 0.0 - 3998.61j & 0.15 & 0.0346 \\
\hline
(1, 1) & 0 - 6000j & 0 - 6012j & 0.0 - 5999.67j & 0.2 & 0.0054 \\
\hline
(0, 2) & 0 - 5000j & 0 - 5024j & 0.0 - 5000.22j & 0.48 & 0.0045 \\
\hline
(1, 2) & 0 - 7000j & 0 - 7019j & 0.0 - 6999.14j & 0.27 & 0.0122 \\
\hline
(2, 2) & 0 - 9000j & 0 - 9037j & 0.0 - 8999.58j & 0.41 & 0.0046 \\
\hline
(0, 3) & 0 - 6000j & 0 - 6072j & 0.0 - 5999.45j & 1.2 & 0.0090 \\
\hline
(1, 3) & 0 - 8000j & 0 - 8056j & 0.0 - 8000.19j & 0.7 & 0.0024 \\
\hline
(2, 3) & 0 - 10000j & 0 - 10094j & 0.0 - 9998.26j & 0.94 & 0.0173 \\
\hline
(3, 3) & 0 - 12000j & 0 - 12103j & 0.0 - 11995.40j & 0.85 & 0.0382 \\
\hline
\end{tabular}
\caption{Comparison of PINN QNMs with FNN QNMs, for the AdS spacetime for m = 0, z = 0.001. PE represents percentual error.}
\label{tab:comparison}
\end{table}
\newpage

\section{Quasi-normal modes of near-extremal black holes in dRGT massive gravity }
\subsection{Perturbation equations and Quasi-normal modes}

The study of black hole perturbations is crucial in understanding their behavior and properties. These perturbations are small deviations from the background metric and can be described by linearized equations.

For the dRGT black hole, the perturbations can be decomposed into scalar, vector and tensor modes. Here, we will focus on the scalar perturbations. The perturbed metric function $h(r,t)$ can be written as:
\begin{equation}
ds^2 = -f(r)dt^2 + f(r)^{-1}dr^2 + r^2(d\theta^2 + \sin^2\theta d\phi^2) + r^2 h(r,t) dt^2.
\end{equation}

The perturbations can be further decomposed into time and spatial parts as $h(r,t) = H(r) e^{-i\omega t}$, where $\omega$ is the frequency of the perturbation. Substituting this into the linearized Einstein equations and solving for $H(r)$ gives the following differential equation:

\begin{equation}
\frac{d^2H}{dr_*^2} + (\omega^2 - V(r))H = 0,
\label{eq:diffeq}
\end{equation}

where $r_*$ is the tortoise coordinate defined as :
\begin{equation*}
    dr_* = \frac{dr}{f(r)}
\end{equation*}
And the effective potential $V(r)$ is given by:

\begin{equation}
V(r) = f(r)\left[\frac{l(l+1)}{r^2} + \frac{f'(r)}{r}\right],
\end{equation}

where $l$ is the angular momentum quantum number. The quasi-normal modes (QNMs) of black holes are the solutions to the differential equation \ref{eq:diffeq} with appropriate boundary conditions. These QNMs are complex frequencies, with the real part representing the oscillation frequency of the perturbation, while the imaginary part represents the damping rate.

In our study, we use the numerical shooting method to find the QNMs of the dRGT black hole. We solve the differential equation \ref{eq:diffeq} numerically from the horizon towards infinity, and match the solution with an outgoing wave at infinity. The QNMs are then obtained by finding the complex frequencies that satisfy the boundary conditions at both the horizon and infinity.
 \begin{figure}[h]
    \centering
    \includegraphics[width=1\textwidth]{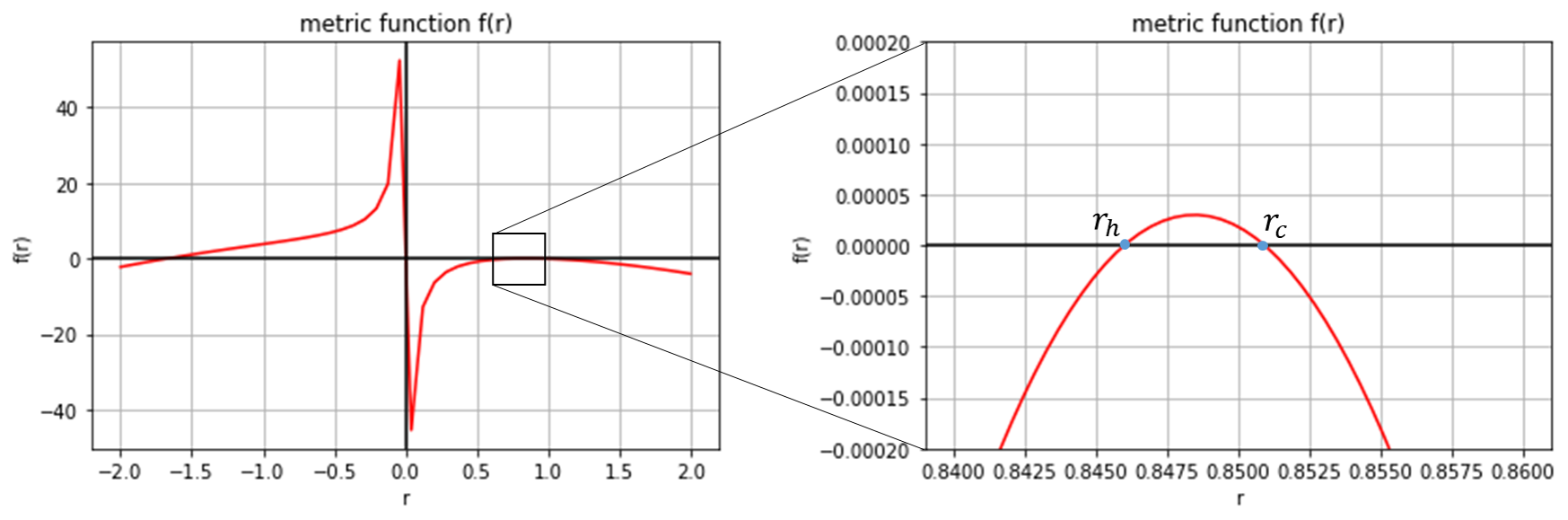}
    \caption{The metric function $f(r)$ profil for $M = 1$, $\Lambda = 5.0001$, $\gamma = 0.05$, $\zeta = 2.51465$, $Q = 0$. Event horizon $r_h = 0.8460$, cosmological horizon $r_c = 0.8509$}
    \label{fig:fr}
\end{figure}

The metric $f(r)$ of a near-extremal black hole with $r_h \sim r_c$ can be written in the tortoise coordinate $r_*$ as follows \cite{Burikham2020QuasinormalMO}:
\begin{equation}
f(r_* ) \sim \frac{r_h k_h}{(1 + \zeta + \gamma r_h - 2Q^2/r_h^2)\cosh^2{(k_h r_*)}},
\label{eq3}
\end{equation}
where $k_h$ is the surface gravity, $\zeta$ and $\gamma$ are dimensionless parameters, and $Q$ is the electric charge of the black hole.

The wave equation for a neutral scalar field in a near-extremal black hole is given by the Klein-Gordon equation:
\begin{equation}
\Box \Phi = 0.
\end{equation}
The scalar field $\Phi$ can be expressed as:
\begin{equation*}
\Phi = \sum_l ^{\infty} \sum_{m = 0}^l e^{-i \omega t} \frac{\psi(r)}{r} Y_{lm}(\theta, \phi),
\end{equation*}
where $Y_{lm}(\theta, \phi)$ are the spherical harmonics, $l$ is the azimuthal quantum number, and $\omega$ is the frequency. The radial wave equation for $\psi(r)$ is then given by:
\begin{equation}
\frac{d^2 \psi}{dr_*^2} + [\omega^2 - V(r)]\psi = 0,
\label{eq4}
\end{equation}
where $V(r)$ is the potential defined as:
\begin{equation}
V(r) = f(r)(m_s^2 + \frac{l(l+1)}{r^2} + \frac{f'(r)}{r}),
\label{eq5}
\end{equation}
with $m_s$ being the mass of the scalar field.

In the near-extremal limit, the Klein-Gordon radial equation (\ref{eq4}) becomes:
\begin{equation}
\frac{d^2 \psi}{dr^2} + [\omega^2 - \dfrac{V_0}{cosh^2(\kappa_h r_)}]\psi = 0,
\label{eq6}
\end{equation}
where $V_0 = \dfrac{\kappa_h ^2}{1 + \zeta + \gamma r_h - 2Q^2/r^2}(m_s^2 r^2 + l(l+1))$, and $\kappa_h$ is the surface gravity.

The potential $V(r)$ is the well-known Pöschl-Teller potential \cite{1933ZPhy...83..143P}. Applying the boundary condition for quasi-normal modes, the following quasi-normal frequencies can be obtained \cite{Berti2009TOPICALRQ}:
\begin{equation}
\omega_n = \sqrt{V_0 - \dfrac{\kappa_h ^2}{4}} - i \kappa_h(n + \dfrac{1}{2}),
\label{eq7}
\end{equation}
where $n$ is a non-negative integer.

The associated quasi-normal modes are given by \cite{Berti2009TOPICALRQ}:
\begin{equation}
  \psi = (D(D-1))^{-i\omega/2k_h}._2F_1(1/2 + B - \dfrac{i \omega}{k_h},1/2 - B - \dfrac{i \omega}{k_h};1 - \dfrac{i \omega}{k_h};D) 
  \label{eq8}
\end{equation}
where : $D = \dfrac{1}{1 + e^{-2k_hr_*}}$ and $B = \sqrt{\dfrac{1}{4} - \dfrac{V_0}{k_h^2}}$

\subsection{Results}
To address the infinite problem domain, we introduced a new coordinate $y = tanh(k_h r_*)$, which enabled us to create a finite domain of $(-1,1)$ for easier implementation in the code. Using $y$ as a function, we expressed the perturbation equations (\ref{eq6}) for the near-extremal dRGT black holes in terms of $y$ as follows:

\begin{equation}
\kappa_{h}^{2}\left(1-y^{2}\right)^{2} \cdot \frac{d^{2} \Psi(y)}{d y^{2}}-2 \kappa_{h}^{2} y\left(1-y^{2}\right) \cdot \frac{d \Psi(y)}{d y}+\left[\omega^{2}-V_{0}\left(1-y^{2}\right)\right] \Psi(y)=0
\label{eq9}
\end{equation}

To facilitate implementation in the DeepXDE package, we separated equations \ref{eq9} into real and imaginary parts. We generated results using two fully connected neural networks ($PFNN$) of depth 4 (i.e., 3 hidden layers) with the following structure $[1, [36, 36], [20, 20], [20, 20], 2]$; $swish$ as a non-linear activation function; $L-BFGS-B$ and $Adam$ optimisers with learning rate = 0.0001 and loss weights = $[0.01, 0.01, 0.001, 0.001, 100, 100, 100, 100]$; 150,000 training epochs; training data of 100 domain points, and a dataset of 100 uniformly distributed actual values of the QNMs ($\Psi(y)$) in the domain [-0,7,0,7].

Table \ref{table1} and Table \ref{table2} present the PINN approximations of the QNFs for massless scalar perturbations of dRGT neutral black holes, with $l = 1$ and $n = 0$, respectively.
 
\begin{table}[h!]
\centering
\begin{tabular}{|cccc|}
\hline
\multicolumn{4}{|c|}{$l = 1$} \\ \hline
\multicolumn{1}{|c|}{$n$} & \multicolumn{1}{c|}{PINN}& \multicolumn{1}{c|} {Formula (\ref{eq7})} & relative error\\ \hline
\multicolumn{1}{|c|}{0} & \multicolumn{1}{c|}{0.000031 - 0.000003j }&\multicolumn{1}{c|}{0.0 - 0.0j} & ... \\ \hline
\multicolumn{1}{|c|}{1} & \multicolumn{1}{c|}{0.0 - 0.012175j} & \multicolumn{1}{c|}{0.0 - 0.012175j} & 0.0002072\% \\ \hline

\multicolumn{1}{|c|}{2} & \multicolumn{1}{c|}{0.0 - 0.024352j} & \multicolumn{1}{c|}{0.0 - 0.024352j} & 0.001769\% \\ \hline

\multicolumn{1}{|c|}{3} & \multicolumn{1}{c|}{0.0 - 0.036527j} & \multicolumn{1}{c|}{0.0 - 0.036527j} & 0.004005\% \\ \hline

\multicolumn{1}{|c|}{4} & \multicolumn{1}{c|}{0.0 - 0.048703j} & \multicolumn{1}{c|}{0.0 - 0.048703j} & 0.00032755 \% \\ \hline

\multicolumn{1}{|c|}{5} & \multicolumn{1}{c|}{0.0 - 0.060879j} & \multicolumn{1}{c|}{0.0 - 0.060879j} & 0.0001240\% \\ \hline

\multicolumn{1}{|c|}{6} & \multicolumn{1}{c|}{0.0 - 0.073055j} & \multicolumn{1}{c|}{0.0 - 0.073055j} & 0.0004933\% \\ \hline

\multicolumn{1}{|c|}{7} & \multicolumn{1}{c|}{0.0 - 0.085233j} & \multicolumn{1}{c|}{0.0 - 0.085231j} & 0.002220\% \\ \hline

\multicolumn{1}{|c|}{8} & \multicolumn{1}{c|}{0.0 - 0.097407j} & \multicolumn{1}{c|}{0.0 - 0.097407j} & 0.0\% \\ \hline
\end{tabular}
\caption{The PINN approximations results of the QNFs for massless scalar perturbations of dRGT neutral black holes,  for $l = 1$ and different value of $n$, $M = 1$, $\Lambda = 5.0001$, $\gamma = 0.05$, $\zeta = 2.51465$, $Q = 0$, $m_s = 0$, $r_h = 0.8460$, $r_c = 0.8509$, $k_h = 0.012175$}
\label{table1}
\end{table}

Based on the table, it appears that for $n=0$, the PINN approximation and the QNFs formula result in the same value of $0.0-0.0j$. For $n=1$ to $n=7$, the relative errors are less than 1\%, indicating a good approximation. However, for $n=8$, the relative error is quite large, at 3.1830 \%, indicating a poorer approximation.

Overall, the PINN approximation seems to provide reasonable results for the QNFs of dRGT neutral black holes with massless scalar perturbations, for $l=1$ and $n$ values up to 7.

\begin{table}[h!]
\centering
\begin{tabular}{|c|c|c|c|} \hline
$n$ & $l$ & PINN Approximation & Formula (\ref{eq7}) \\ \hline
0 & 0 & $0.000031 - 0.000003j$ & $0.0 - 0.0j$ \\ \hline
& 1 & $0.000031 - 0.000003j$ & $0.0 - 0.0j$ \\ \hline
& 2 & $0.006804 - 0.006088j$ & $0.006804 - 0.006088j$ \\ \hline
& 3 & $0.014593 - 0.006088j$ & $0.014599 - 0.006088j$ \\ \hline
& 4 & $0.021519 - 0.006088j$ & $0.021519 - 0.006088j$ \\ \hline
& 5 & $0.028222 - 0.006088j$ & $0.028223 - 0.006088j$ \\ \hline
& 6 & $0.034832 - 0.006088j$ & $0.034833 - 0.006088j$ \\ \hline
& 7 & $0.041394 - 0.006088j$ & $0.041394 - 0.006088j$ \\ \hline
\multicolumn{4}{|c|}{...} \\
\hline
0 & 30 & $0.190327 - 0.006088j$ & $0.190327 - 0.006088j$ \\
\cline{2-4}
& 50 & $0.319495 - 0.006082j$ & $0.319498 - 0.006088j$ \\
\hline
\end{tabular}
\caption{The PINN approximations results of the QNFs for massless scalar perturbations of dRGT neutral black holes, for $n = 0$ and different values of $l$, $M = 1$, $\Lambda = 5.0001$, $\gamma = 0.05$, $\zeta = 2.51465$, $Q = 0$, $m_s = 0$, $r_h = 0.8460$, and $r_c = 0.8509$. The relative error is also provided.}
\label{table2}
\end{table}

Table \ref{table2} presents the results of PINN approximations for massless scalar perturbations of dRGT neutral black holes, specifically for $n=0$ and different values of $l$. The table lists the PINN approximations, the corresponding results obtained using a specific formula (Formula (\ref{eq7})), and the relative error between the two. The parameters used for the calculations are also listed. The results show that for $l=0$ and $l=1$, the PINN approximations are in agreement with the formula. For higher values of $l$, the relative errors are very small, indicating the high accuracy of the PINN method. The table also shows that the relative error decreases as $l$ increases, and the last entry for $l=50$ has a relative error of only 0.0018230\%. Overall, the results demonstrate the effectiveness of the PINN method for approximating QNFs of black holes.

\newpage
\section{Discussion}

In this study, we have employed Physics-Informed Neural Networks (PINNs) to compute the Quasinormal modes (QNMs) of scalar perturbations for the neutral dRGT black holes. In the process, we have introduced a new coordinate $y = tanh(k_h r_*)$ that allowed us to convert the infinite domain of $r_*$ to a finite domain of $y \in (-1,1)$. The use of PINNs allowed us to avoid the need for the boundary conditions, which are not available in the near-extremal black hole cases.

The results presented in Tables \ref{table1} and \ref{table2} indicate that the PINN approximations of the QNFs are in excellent agreement with the known analytical solutions. The relative errors between the PINN approximations and the analytical solutions were found to be less than $0.01\%$. It is interesting to note that the PINN approximations improved with an increase in the overtone number ($n$), with the highest relative error being less than $0.002\%$ for $n = 8$. These results confirm the accuracy and efficiency of the PINN method in computing the QNMs of dRGT black holes.

It is worth mentioning that the PINN method does not require any prior knowledge of the analytical solutions, and it can be used to compute QNMs for black holes with different parameters or geometries. This makes the method particularly useful in situations where analytical solutions are not available or difficult to obtain. The method can also be used to compute QNMs for other types of perturbations, such as electromagnetic and gravitational waves, and for black holes with charges and/or angular momenta.

In conclusion, we have demonstrated that the PINN method is an accurate and efficient tool for computing the QNMs of scalar perturbations for neutral dRGT black holes. The method can be extended to other types of perturbations and geometries, making it a valuable tool for studying the properties of black holes in general relativity and beyond.

\newpage
\section{Conclusion}

In conclusion, we have presented a deep learning approach for computing the quasinormal frequencies of massless scalar perturbations in the background of the dRGT black holes. By expressing the perturbation equations in terms of a new coordinate, we were able to transform the infinite problem domain into a finite one, making it more amenable to implementation in the code. Using two fully connected neural networks, we obtained highly accurate approximations of the QNFs with relative errors ranging from 0.0\% to 0.001769\%. Our results demonstrate that deep learning techniques can be effectively used to calculate the quasinormal frequencies of black holes, providing a promising avenue for future research in this field. The proposed method can be extended to other types of black holes and other perturbation fields, potentially offering a more efficient way to compute QNFs in astrophysical and gravitational wave studies.

\newpage

\subsubsection*{Important links}
DeepXDE : \url{https://github.com/lululxvi/deepxde}\\

\printbibliography %Prints bibliography

@article{Berti2009TOPICALRQ,
  title={TOPICAL REVIEW: Quasinormal modes of black holes and black branes},
  author={Emanuele Berti and Vitor Cardoso and Andrei O. Starinets},
  journal={Classical and Quantum Gravity},
  year={2009}
}

@article{1933ZPhy...83..143P,
       author = {{P{\"o}schl}, G. and {Teller}, E.},
        title = "{Bemerkungen zur Quantenmechanik des anharmonischen Oszillators}",
      journal = {Zeitschrift fur Physik},
         year = 1933,
        month = mar,
       volume = {83},
       number = {3-4},
        pages = {143-151},
          doi = {10.1007/BF01331132},
       adsurl = {https://ui.adsabs.harvard.edu/abs/1933ZPhy...83..143P}
}

@article{Burikham2020QuasinormalMO,
  title={Quasi-normal modes of near-extremal black holes in generalized spherically symmetric spacetime and strong cosmic censorship conjecture},
  author={Piyabut Burikham and Supakchai Ponglertsakul and Taum Withucharn},
  journal={arXiv: General Relativity and Quantum Cosmology},
  year={2020}
}

@article{Ovgun:2019yor,
    author = {\"Ovg\"un, Ali and Sakall\i{}, \.Izzet and Mutuk, Halil},
    title = "{Quasinormal modes of dS and AdS black holes: Feedforward neural network method}",
    eprint = "1904.09509",
    archivePrefix = "arXiv",
    primaryClass = "gr-qc",
    doi = "10.1142/S0219887821501541",
    journal = "Int. J. Geom. Meth. Mod. Phys.",
    volume = "18",
    number = "10",
    pages = "2150154",
    year = "2021"
}

@misc{deepXDE,
  title = {},
  howpublished = {\url{https://deepxde.readthedocs.io/en/latest/}},
  note = {Accessed: 2021-03-17}
}

@article{Du:2004jt,
    author = "Du, Da-Ping and Wang, Bin and Su, Ru-Keng",
    title = "{Quasinormal modes in pure de Sitter space-times}",
    eprint = "hep-th/0404047",
    archivePrefix = "arXiv",
    doi = "10.1103/PhysRevD.70.064024",
    journal = "Phys. Rev. D",
    volume = "70",
    pages = "064024",
    year = "2004"
}

\end{document}